\documentclass[12pt]{article}

\usepackage[english]{babel}
\usepackage{float} 
\usepackage{ulem}
\usepackage{amssymb,multirow}
\usepackage{subfigure}
\newtheorem{theorem}{Theorem}[section]
\newtheorem{proposition}{Proposition}[section]
\newtheorem{assumption}{Assumption}[section]
\newtheorem{remark}{Remark}[section]
\newcommand{\dd}{\mathrm{d}}

\usepackage[letterpaper,top=2cm,bottom=2cm,left=3cm,right=3cm,marginparwidth=1.75cm]{geometry}

\usepackage{amsmath}
\usepackage{graphicx}
\usepackage[colorlinks=true, allcolors=blue]{hyperref}
\usepackage{authblk}
\usepackage[authoryear]{natbib}

\title{Non-Parametric Estimation Techniques of Factor Copula Model using Proxies}
\author[1]{Bahareh Ghanbari\thanks{Corresponding author. Email: \texttt{Bahareh.Ghanbari@rmit.edu.au}}}
\author[2]{Pavel Krupskiy}
\author[1]{Laleh Tafakori}
\author[1]{Yan Wang}

\affil[1]{Department of Mathematical Sciences, RMIT University, Melbourne, Australia.}
\affil[2]{School of Mathematics and Statistics, The University of Melbourne, Melbourne, Australia.}

\begin{document}

\maketitle

\begin{abstract}
Parametric factor copula models typically work well in modeling multivariate dependencies due to their flexibility and ability to capture complex dependency structures. However, accurately estimating the linking copulas within these models remains challenging, especially when working with high-dimensional data. This paper proposes a novel approach for estimating linking copulas based on a non-parametric kernel estimator. Unlike conventional parametric methods, our approach utilizes the flexibility of kernel density estimation to capture the underlying dependencies more accurately, particularly in scenarios where the underlying copula structure is complex or unknown. We show that the proposed estimator is consistent under mild conditions and demonstrate its effectiveness through extensive simulation studies.
Our findings suggest that the proposed approach offers a promising avenue for modeling multivariate dependencies, particularly in applications requiring robust and efficient estimation of copula-based models. 

 
\end{abstract}

\section{Introduction}
Factor copula models provide a flexible yet concise framework, assuming that variables become independent when conditioned on one or more unobserved common factors. The general class of factor copula models proposed by \cite{Krupskii2013}, which is an extension of classical Gaussian factor models, explains the dependence structure of high dimensional variables in terms of a few latent variables. \cite{Krupskii2015} extended their approach to model dependence when the observed variables belong to several non-overlapping groups with homogeneous dependence in each group. 

Factor copulas have demonstrated their ability to capture both the correlations and dependence in the joint tails, thanks to the flexibility of copula functions. Since introducing the Gaussian factor copula model by \cite{Hull2004}, factor copulas have been further developed to account for various data characteristics. For instance, \cite{Chen2015} employed factor copulas to examine the mortality dependence of multiple populations and factor copulas have also been used to study the behavior dependence of item responses in \cite{Nikoloulopoulos2015}. In \cite{Oh2017}, it is shown that a linear factor copula model, as a particular case of factor models, can be used to capture dependence between economic variables while the properties of these models and their extreme-value limits are studied by \cite{Krupskii2018}. 
Financial time series can significantly benefit from the factor copula models as they provide a superior fit compared to truncated vines (\cite{Brechmann2012}) and classical Gaussian factor models, especially when utilizing stock returns from the same industry sector.  

In parametric factor copula models, the choice of bivariate linking copulas is of interest to define appropriately the dependence structure between observed and latent variables. To deal with this, various diagnostic tools have been proposed; see \cite{Krupskii2017}, which used bivariate normal score plots of the observed data or measures of tail asymmetry. Bayesian inference was used by \cite{Nguyen2019} as an alternative approach, but it is computationally demanding. \cite{Krupskii2022} showed how the use of proxies for estimating latent variables in factor copula models can aid in selecting bivariate copula families, while \cite{Fan2023} extended their work to bi-factor and other structured $p$-factor copulas using two-stage proxies. 

Parametric models for linking copulas are frequently utilized, but they can lead to incorrect specifications and inconsistencies in estimators. The most common parametric estimation methods include maximum likelihood employing a two-stage procedure, as demonstrated in \cite{Ko2019}, or penalized likelihood, as illustrated in \cite{Qu2012} and more recently applied to Archimax copulas according to \cite{Cha2020}. However, these methods can be restricted by the number of parameters and may only be suitable for copulas that conform to the widely used parametric families.
In contrast to traditional estimation techniques, non-parametric estimation does not assume that data are drawn from a known distribution. Instead, non-parametric models determine the model structure from the underlying data. These methods provide a more flexible alternative with several different approaches, such as deep learning theory \cite{Liu2024}, kernel estimators \cite{Nagler2016, Wang2023, Modak2023}, B-spline estimators \cite{Kau2013,Kirkby2023}, Bernstein polynomials \cite{Bou2010,Bou2013,kiriliouk2018estimator}, linear wavelet estimators \cite{Gen2009,Ghanbari2019}, nonlinear wavelet estimators \cite{Autin2010,Ghanbari2023}, or Legendre multiwavelet estimators \cite{Chat2017}.

Non-parametric kernel methods are a class of statistical techniques widely employed in data analysis and machine learning, offering flexible and versatile solutions particularly suitable for situations where underlying data distributions are not well-known or may deviate from parametric assumptions. \cite{Jin2021} proposed an improved kernel density estimation based on $L_2$ regularization term. \cite{Ye2021} studied a novel kernel density estimator to determine the expectation of an estimated PDF using an ensemble of data depending on unbiased cross-validation. These methods avoid explicit functional form assumptions and instead rely on local information around data points to make inferences. Central to these approaches are kernel functions, which efficiently capture local patterns and relationships within the data. Typical applications of non-parametric kernel methods include density estimation, regression analysis, and classification tasks. In kernel density estimation, these methods allow for constructing smooth and continuous probability density functions without imposing rigid distributional assumptions. Moreover, non-parametric kernel regression methods enable the modeling of complex relationships between variables by incorporating local weighted averages. The adaptability and resilience to diverse data structures make non-parametric kernel methods valuable in various scientific disciplines, offering practical solutions for analyzing and understanding complex datasets without requiring restrictive parametric assumptions. 

The use of kernel methods in analyzing high-dimensional datasets has gained increasing attention from researchers and practitioners due to the complex datasets characterized by numerous variables. Traditional parametric methods often need help with challenges such as the curse of dimensionality and the risk of model overfitting in high-dimensional spaces \cite{Scott1992, Bellman2003}. Kernel methods provide a valuable alternative by implicitly mapping the data into a higher-dimensional space using kernel functions. This transformation captures intricate relationships among variables, facilitating the extraction of meaningful patterns in high-dimensional data. For example, Support Vector Machines (SVMs) equipped with kernel functions have been proven effective in classification tasks where the number of features significantly exceeds the number of observations \cite{Sanchez2003}. Additionally, kernel principal component analysis enables dimensionality reduction while preserving the non-linear structures inherent in the data. Recent research has also focused on developing novel kernel functions specifically tailored to address the challenges posed by high-dimensional datasets, enhancing the adaptability and performance of kernel methods in this context \cite{Liu2021}.

Although the kernel estimators are widely utilized for estimating general densities, to our knowledge, our work is the first to apply these estimators to estimate the factor copula density. Inspired by the technique proposed by \cite{Nagler2016} for linking copulas within vine copula models to overcome the curse of dimensionality, we introduce a novel non-parametric kernel-based method for estimating linking copulas in one-factor copula models. Our analysis focuses explicitly on its computational complexity in the presence of a single latent variable and studies the asymptotic properties of the estimator which uses proxy methods to estimate the unobserved latent variable. Our method is beneficial when the factor copula model and the linking copulas must be correctly specified. This alternative approach enables the selection of suitable linking copulas necessary for modeling dependency when the common factor drives the dependence among observed variables. 

The remainder of this paper is structured as follows. In Section 2 we review the class of one-factor copula models, introduce the proposed kernel estimator, and study its asymptotic properties. In Section 3, we evaluate the performance of the proposed method through an extensive simulation study. In Section 4,  we use the proposed methodology for the analysis of financial stock returns data, and Section 5 concludes the paper with a discussion of the findings and potential directions for future research. Lastly, the proof of the theoretical results is provided in the Appendix.

\section{Methodology}\label{sec2}
Both parsimonious and flexible-factor copula models can handle the dependence between the observed variables based on a few unobserved (latent) variables. These models are particularly useful when tail asymmetry or dependence is observed in multivariate data, so the standard models based on the multivariate normality assumption are unsuitable. 
Recently, \cite{Oh2017, Krupskii2013} proposed factor copula models, which overcome some disadvantages of traditional copulas and are more flexible than the classical Gaussian factor models. The first approach combines the class of dynamic factor models commonly used in time series analysis with arbitrary marginal distributions. However, this approach restricts the choice of copula functions to certain extensions of elliptical distributions, such as the Student's-$t$ and skew Student's-$t$ copulas. 

Alternatively, the second model assumes that the dependence among observed variables can be captured using a multivariate factor copula model. In this framework, it is assumed that the dependence structure arises from a set of $p$ latent factors $V_1,\ldots, V_p$.  A common approach is to transform the observed variables $X_1,\ldots, X_d$ into uniform random variables $\boldsymbol{U} = (U_1,\ldots, U_d)$, where each $U_i\sim U(0,1)$. The transformed variables $U_1,\ldots, U_d$ satisfy $U_i=F_i(X_i)$, where $F_i$ is the marginal cumulative distribution function (CDF) of  $X_i$. 
The dependence among these variables is captured by a copula function $C(u_1,\ldots, u_d)$, which allows for a flexible representation of a complex dependence structure.

In this paper, we focus on the one-factor copula model, where a single latent factor drives the dependence among observable variables. This structure strikes a balance between simplicity and flexibility, making it particularly suitable for high-dimensional datasets. Specifically, we estimate the one-factor copula model non-parametrically using a kernel estimation technique, allowing for greater flexibility in capturing complex dependence structures without restrictive parametric assumptions.

\subsection{One-factor copula models}\label{b}
The one-factor copula model provides a flexible framework for capturing dependence among multiple variables while maintaining computational tractability \cite{Krupskii2013}. In this model, we consider $d$ observable variables $U_1,\ldots, U_d$, each following a uniform distribution on $(0,1)$. These variables are assumed to be conditionally independent given an underlying latent factor $V_0$, which follows a uniform distribution: $V_0 \sim U(0,1)$. 

To describe the dependence structure between each $U_i$ and the latent factor $V_0$, we introduce the bivariate copula density $c_{U_j, V_0}(u_j,v_0)$ and their CDFs $C_{U_j, V_0}(u_j,v_0)$. Under the one-factor framework, the joint CDF of $\boldsymbol{U} = (U_1,\ldots, U_d)$ can be expressed as: 
\begin{equation}
 C_{\boldsymbol{U}}(u_1,\ldots,u_d) = \int_0^1 \prod_{j=1}^d C_{U_j|V_0}(u_j|v_0) \dd v_0,   
\end{equation}
Here $C_{U_j|V_0}$ is defined as the conditional CDF of $U_j$ given $V_0$. 

Let us denote the partial derivative $ \frac{\partial}{\partial v_0} C_{U_j,V_0}(u_j,v_0) := C_{U_j|V_0}(u_j|v_0)$. Note that $\frac{\partial}{\partial u_j} C_{U_j|V_0}(u_j|v_0) = \frac{\partial^2}{\partial u_j \partial v_0} C_{U_j,V_0}(u_j,v_0) = c_{U_j,V_0}(u_j,v_0)$.
As shown in \cite{Krupskii2013}, the copula density for the one-factor copula model is:
\begin{equation}\label{fcd}
 c_{\boldsymbol{U}}(u_1,\ldots,u_d) = \frac{\partial^d C(u_1,\ldots,u_d)}{\partial u_1\ldots\partial u_d} = \int_0^1 \prod_{j=1}^d c_{U_j,V_0}(u_j,v_0) \dd v_0.   
\end{equation}

The factor copula models have become increasingly popular in recent years due to their ability to account for dependencies among the observed variables. However,  these dependencies are mostly determined by the unobserved common factors, referred to as latent variables. Consequently, estimating these latent variables can help select appropriate bivariate linking copulas or accurately estimate them. 

To achieve that, \cite{Krupskii2022} proposed a new proxy method that uses the unweighted averages computed from the observed variables $U_1, \ldots, U_d$. Denote $Z_j= \Phi^{-1} (U_j)$ and $W = \Phi^{-1} (V_0)$.  To construct the proxy for $W$, we first compute the average of the transformed variables as follows:
\begin{eqnarray}\label{vhat}
    \Bar{Z}_d = \frac{1}{d} \sum_{j=1}^d \Phi^{-1} (U_j).
\end{eqnarray}
Then, we compute $\hat{F}(\Bar{Z}_d)$, where $\hat{F}$ is the empirical CDF of $\Bar{Z}_d$. This step transforms $\Bar{Z}_d$ to the uniform scale. Finally the proxy for $W$ is obtained by applying the inverse normal CDF,  $\hat{W} =\Phi^{-1}(\hat{F}(\Bar{Z}_d))$. This transformation ensures that $\hat W$ is approximately normally distributed. To use the proxy method in \cite{Krupskii2022}, it is required that the bivariate linking copulas have monotonic dependence (in particular, stochastically increasing copulas satisfy this requirement), implying that the observed variables are monotonically related to the latent variable, which is not a very restrictive assumption in many applications.

\subsection{Proposed non-parametric estimator}\label{a}

We develop the non-parametric estimator of the one-factor copula density, which involves the kernel estimation of the bivariate linking copulas. The primary challenge here is estimating the linking copulas in the presence of a latent variable. To address this, we initially apply the proxy method introduced by \cite{Krupskii2022} to estimate the latent variable, followed by the kernel method to estimate the linking copulas. 

The idea behind estimating the factor copula density function in \eqref{fcd} is to estimate each bivariate linking copula density individually while taking into consideration that there is a key distinction between this approach and typical kernel estimation techniques that have been extensively studied in the literature. The key distinction of this approach compared to conventional kernel estimation techniques lies in its consideration of a latent variable. Unlike standard kernel estimation methods, which typically estimate density functions directly from observable data, the proposed procedure accounts for the dependence structure induced by an unobservable (latent) factor. We employ a step-wise estimation procedure to systematically detail our non-parametric estimator's construction.

\begin{itemize}
    \item \textbf{Transform observed data into uniform variables $U_i$ :} Assume that $\boldsymbol{X} = (X_1,\ldots,X_d)$ is a random vector with the joint cdf $F$ and marginal cdfs $F_1,\ldots,F_d$. We obtain the estimates $\hat{F}_1,\ldots,\hat{F}_d$ of the respective marginal CDFs based on all observations ($X_1^{(i)},\ldots, X_d^{(i)}$) for $i= 1,\ldots,n$. Let $\hat{F}_j$ be the kernel estimator of $F_j$, $j=1,\ldots,d$, which is defined in \cite{Nagler2017}: 
\begin{equation}\label{standard}
  \hat{F}_j(x) = \frac{1}{n} \sum_{i=1}^n J\left(\frac{X_j^{(i)}-x}{b_n}\right),
\end{equation}
where $J(x) = \int_{-\infty}^x K(s) ds$ for all $x\in \mathbb{R}$, and $K$ is a kernel function with the bandwidth $b_n>0$ that satisfy the following conditions:
\begin{itemize}
\item  C1: $K$ is a symmetric density function supported on $[-1,1]$ and has continuous first-order derivative. 
\item  C2: $b_n \downarrow 0$ and $nb_n^2 \rightarrow \infty$ as $n \to \infty$.
\end{itemize}
To estimate the latent variable and linking copula densities, we need data on $U(0,1)$ scale, i.e., the random vector $\boldsymbol{U} = (F_1(X_1),\ldots, F_d(X_d))$. In practice, we do not have access to observations from this vector; instead, we can use pseudo-observations $\hat{\boldsymbol{U}}^{(i)} = (\hat{U}_1^{(i)},\ldots,\hat{U}_{d}^{(i)})$ by replacing $F_1,\ldots, F_d$ with their estimators:
\begin{equation}\label{psed}
  (\hat{U}_1^{(i)},\ldots,\hat{U}_{d}^{(i)}) = (\hat{F}_1(X_1^{(i)}),\ldots, \hat{F}_d(X_d^{(i)})),~~~~~i=1,\ldots,n. 
\end{equation}
   \item \textbf{Estimating the latent factor $V_0$:}
    Utilizing the full dataset in \eqref{psed}  with $d$ variables, we first compute the proxy as defined in \eqref{vhat}. To do that, we need to compute $\hat{F}(\Bar{Z}_d)$ where $\hat{F}$ is the empirical CDF of $\Bar{Z}_d$. We then use this proxy (the estimated latent variable) to estimate the density of the copula linking $K$ variables denoted $c_{1:K}$ (without loss of generality, we can assume these are the first $K$ variables).

Copula densities are supported on the unit hypercube, which necessitates careful consideration during estimation. Only a few kernel estimators are suitable for addressing bias and consistency issues that arise at the boundaries of the support. To overcome these challenges and be able to estimate the $K$-dimensional marginal copula density $c_{1:K}$, represented as the product of the $K$ bivariate linking copulas $c_{U_j, V_0}$ in the presence of a latent variable, we employ the techniques of the transformation estimator presented in \cite{Nagler2017}, which transforms the data into standard normal margins, thereby enabling unbounded support. Assume $\Phi$ is the standard Gaussian distribution function, while $\phi$ and $\Phi^{-1}$ are its density and quantile functions, respectively. According to \cite{Sklar1959}, 
The joint density $f$ of variables $(Z_1^{(i)}, Z_2^{(i)})=(\Phi^{-1}(U_1^{(i)}),\Phi^{-1}(U_2^{(i)}))$, $i=1,\ldots, n$ can be expressed in terms of the respective copula density $c$ and marginal densities as:
\begin{equation}
  f(z_1,z_2) = c(\Phi(z_1), \Phi(z_2))\phi(z_1)\phi(z_2), \quad z_1,z_2\in \mathbb{R}.  
\end{equation}
Changing variables $u_j = \Phi(z_j),$ $j=1,2$, we get
\begin{equation}\label{c12}
  c(u_1, u_2) = \frac{f(\Phi^{-1}(u_1), \Phi^{-1}(u_2))}{\phi(\Phi^{-1}(u_1))\phi(\Phi^{-1}(u_2))}.  
\end{equation}
Using a kernel estimator of $f(z_1, z_2)$ defined as $$\hat{f}(z_1,z_2) =  \frac{1}{nb_n^2} \sum_{i=1}^n K\left(\frac{Z_1^{(i)}-z_1}{b_n}\right) K\left(\frac{Z_2^{(i)}-z_2}{b_n}\right)\,,$$ we obtain from \eqref{c12} the kernel estimator of the copula density $c(u_1,u_2)$ as follows:
\begin{equation}\label{explode}
  \hat{c}(u_1, u_2) = \frac{\hat{f}(\Phi^{-1}(u_1), \Phi^{-1}(u_2))}{\phi(\Phi^{-1}(u_1))\phi(\Phi^{-1}(u_2))}.  
\end{equation}
The copula density $c(u_1,u_2)$ can take infinite values at the boundaries of the support. To avoid the variance of the estimator defined in \eqref{explode} from exploding, we therefore constrain the estimator to pairs $(u_1, u_2) \in [0.001, 0.999]^2$.
    \item \textbf{Estimating the linking copula densities $c(u_j, v_0)$}:  We follow a few steps to develop the kernel estimator for the bivariate linking copula density function. These steps build on the notation introduced in Section \ref{b} and incrementally refine the estimator to handle practical challenges. First, inspiring \eqref{explode}, we define the estimator of the linking copula density $c_{U_j, V_0}$ provided that $W = \Phi^{-1} (V_0)$ is observed as follows: 
\begin{equation}\label{ctilde}
  \Tilde{c}_{U_j, V_0}(u_j,v_0) = \frac{1}{n} \sum_{i=1}^n \frac{\boldsymbol{K}_{B_n} \left( \Phi^{-1}(u_j) - \Phi^{-1}(U_j^{(i)}) , \Phi^{-1}(v_0) - \Phi^{-1}(V_0^{(i)}) \right)}{\phi(\Phi^{-1}(u_j))\phi(\Phi^{-1}(v_0))}.  
\end{equation}
Here, $\Tilde{c}_{U_j, V_0}(u_j, v_0)$ is a kernel estimator of $c_{U_j, V_0}(u_j, v_0)$ under the idealized scenario where the latent variable $W$ is directly observable. Next, recognizing that the latent variable $W$ is unobservable in practice, we approximate it using the proxy variable defined in \eqref{vhat}. Incorporating this proxy variable, we redefine the estimator as:
\begin{equation}\label{cbar}
  \Bar{c}_{U_j, V_0}(u_j,v_0) = \frac{1}{n} \sum_{i=1}^n \frac{\boldsymbol{K}_{B_n} \left( \Phi^{-1}(u_j) - \Phi^{-1}(U_j^{(i)}) , \Phi^{-1}(v_0) - \hat{W}^{(i)} \right)}{\phi(\Phi^{-1}(u_j))\phi(\Phi^{-1}(v_0))},    
\end{equation}
where $\boldsymbol{K}_{B_n} (\boldsymbol{s}) = \frac{\boldsymbol{K}(B_n^{-1} \boldsymbol{s})}{det(B_n)}$, $\boldsymbol{K}(\boldsymbol{s}) = K(s_1) K(s_2)$, and $B_n \in \mathbb{R}^2$ is positive definite bandwidth matrix $B_n = \begin{pmatrix}
b_{1n} & 0 \\
b_{3n} & b_{2n} 
\end{pmatrix}$ with parameters $b_{1n}, b_{3n}, b_{2n}>0$.

We further refine the estimator to address the practical scenario where pseudo-observations are used to estimate unobservable quantities. Using pseudo-observations defined as in \eqref{psed}, the kernel estimator of the linking copula density $c_{U_j, V_0}(u_j, v_0)$ becomes: 
\begin{equation}\label{cjhat}
  \hat{c}_{U_j, V_0}(u_j,v_0) = \frac{1}{n} \sum_{i=1}^n \frac{\boldsymbol{K}_{B_n} \left( \Phi^{-1}(u_j) - \Phi^{-1}(\hat{U}_j^{(i)}) , \Phi^{-1}(v_0) - \hat{W}^{(i)} \right)}{\phi(\Phi^{-1}(u_j))\phi(\Phi^{-1}(v_0))}.    
\end{equation}
 The estimator incorporates observed data, providing a practical approximation for the linking copula density.
     
    \item \textbf{Estimating of one-factor copula density}:  Following the steps from equation \eqref{ctilde} to \eqref{cjhat}, we propose our novel kernel estimator for the one-factor copula density $c_{1:K}$. This estimator is obtained by integrating the product of the first $K$ estimated linking copula densities over the latent variable:
\begin{equation}\label{chat}
 \hat{c}_{1:K}(u_1,\ldots,u_k)  = \int_0^1 \prod_{j=1}^K \hat{c}_{U_j, V_0}(u_j,v_0) \dd v_0.   
\end{equation}
This integral combines the contributions of all individuals linking copula densities to provide a comprehensive estimate of the one-factor copula density.
\end{itemize}
\subsection{Asymptotic properties}\label{main}
In this section, we establish the asymptotic properties and convergence rate of the estimator proposed in Section \ref{a}, along with the necessary assumptions to ensure consistency. Recall that the variables $U_1,\ldots, U_d$ are assumed to be conditionally independent given an underlying latent factor $V_0$, which follows a uniform distribution: $V_0 \sim U(0,1)$ and also assume $W = \Phi^{-1} (V_0)$ where the $\Phi^{-1}$ is the inverse normal CDF. 

\begin{assumption}\label{ass} Let $\hat{F}$ be the non-parametric estimator of  a distribution function $F$, then the following conditions hold: 
\begin{itemize}
    \item (a) For $j=1,\ldots,d$ and $x_j\in \Omega_{X_j}$ with $\Omega_{X_j}$ be the support of $X_j$, 
\begin{eqnarray*}
 \sup_{x_j\in \Omega_{X_j}}|\hat{F}_j(x_j) - F_j(x_j)| = o_{a.s.}(n^{-r})
 \end{eqnarray*}
\item (b) For $i=1,\ldots,n$
$$\sup_{i}|\hat{W}^{(i)} - W^{(i)}| = O_p\left(\frac{\sqrt{\ln n}}{\sqrt{d}} + \frac{1}{\sqrt{n}}\right)$$
\item (c) For all $(u,v)\in (0,1)^2$
\begin{eqnarray*}
 \tilde{c}_{U_j, V_0} (u_j,v_0) - c_{U_j, V_0} (u_j,v_0) = O_p(n^{-r}).   
\end{eqnarray*}
\end{itemize}
where $r=\frac{p}{2p+2}$, assuming that the copula densities $c_{U_j, V_0}$ are $p \geq 2$ times continuously differentiable on $(0,1)^2$.

\end{assumption}

\begin{remark}
While conditions (a) and (c) are standard and are satisfied for most nonparametric estimators of the marginal CDFs $F_j$ and copula densities $c_{U_j, V_0}$, condition (b) requires stochastically increasing copulas $C_{U_j, V_0}$ with $\int_0^1 \left(\partial C_{U_j|V_0}(u_j|v_0)/\partial v_0\right) \mathrm{d} u_j < -K_L$ for some constant $K_L > 0$ and all $j = 1,\ldots, d$, and it follows from the conditional independence of the observed variables $U_1^{(i)}, \ldots, U_d^{(i)}$ and the uniform upper bound of error for the normal probability approximation \cite{Chernozhukov2017,Koike2021} provided that $\ln^5 n/d \to 0$:
\begin{align*}
\sup_i |\hat{\bar Z}_d^{(i)}  - \bar Z_d^{(i)}| &= O_p(1/\sqrt{n}),\\
\sup_i |\bar Z_d^{(i)} - m_d(V_0^{(i)})| &= O_p(\sqrt{\ln n}/\sqrt{d}),
\end{align*}
where $\bar Z_d^{(i)} = \frac{1}{d} \sum_{j=1}^d U_j^{(i)}$, $\hat{\bar Z}_d^{(i)} = \frac{1}{d} \sum_{j=1}^d \hat U_j^{(i)}$ and $m_d(V_0^{(i)}) = E(\bar Z_d^{(i)}|V_0 = V_0^{(i)})$, where $m_d(v)$ is a monotonically increasing function of $v$ for stochastically increasing linking copulas $C_{U_j,V_0}$. This implies that 
\begin{align*}
\sup_i |\hat{\bar Z}_d^{(i)}  - m_d(V_0^{(i)})| &= O_p(1/\sqrt{n} + \sqrt{\ln n}/\sqrt{d}),\\
\sup_i |\hat F_{0,i} (\hat{\bar Z}_d^{(i)}) - F_{0,i} (\hat{\bar Z}_d^{(i)})|&= O_p(1/\sqrt{n}), \\
\sup_i | F_{0,i} (\hat{\bar Z}_d^{(i)}) - F_{0,i} (m_d(V_0^{(i)}))|&= O_p(1/\sqrt{n} + \sqrt{\ln n}/\sqrt{d}), \\
\sup_i |F_{0,i} (m_d(V_0^{(i)}))  - V_0^{(i)}| &= O_p(1/\sqrt{n} + \sqrt{\ln n}/\sqrt{d}),\\
\end{align*}
and hence
$$\sup_i |\hat V_0^{(i)} - V_0^{(i)}| = O_p(1/\sqrt{n} + \sqrt{\ln n}/\sqrt{d}),$$
where $\hat V_0^{(i)} = \hat F_{0,i}(\hat{\bar Z}_d^{(i)})$, where $F_{0,i}$ and $\hat F_{0,i}$ is the CDF and empirical CDF of $\bar Z_d^{(i)}$, respectively. 
\end{remark}

\begin{proposition}\label{pro1}
Define $b_{e,n} = \sup_{i=1,\ldots,n} |\hat{U}_j^{(i)} - U_j^{(i)}|$. 
Under conditions $C1$ and $C2$ in step $1$ in section \eqref{a}, 
the estimators \eqref{ctilde}, \eqref{cbar} and \eqref{cjhat} satisfy for all $u_j, v_0 \in (0,1)^2$: 
\begin{equation}\label{pro1.eq}
\begin{aligned} \Tilde{c}_{U_j, V_0} (u_j,v_0) - c_{U_j, V_0} (u_j,v_0) &= O_p\left(det(B_n)+ \frac{1}{\sqrt{n~det(B_n)}}\right), \\
 \Bar{c}_{U_j, V_0} (u_j,v_0) - \Tilde{c}_{U_j, V_0} (u_j,v_0) &= O_p\left(\frac{1}{\sqrt{n}} + \frac{\sqrt{\ln n}}{\sqrt{d}}\right), \\
 \hat{c}_{U_j, V_0} (u_j,v_0) - \Bar{c}_{U_j, V_0} (u_j,v_0) &= O_p(b_{e,n}).
\end{aligned}
\end{equation}
\end{proposition}
The proof is given in the Appendix \ref{proof}.

\begin{theorem}\label{theo1}
Let $c_{1:K}(u_1,\ldots,u_k)$ be the one-factor copula density function linking the first $K$ variables and $\hat{c}_{1:K}(u_1,\ldots,u_k)$ be its estimator defined as in \eqref{chat}. Based on Proposition \eqref{pro1} and assuming that there exists a constant $K_0$ such that $c_{U_j,V_0}(u_j,v_0)< K_0$ for all $0 < u_{\min} < u_j < u_{\max} < 1$ ($j = 1, \ldots, K$) and $0\leq v_0\leq 1$, we obtain
\begin{equation}
  \hat{c}_{1:K}(u_1,\ldots,u_k) -  c_{1:K}(u_1,\ldots,u_k) = O_p(n^{-r} + \sqrt{\ln n}/\sqrt{d}).
\end{equation}
\end{theorem}
The proof is deferred to Appendix \ref{proof}.
\begin{remark}
In deriving the results presented above, we consider the full dataset, allowing $d \to \infty$, to estimate the latent variable. However, for the copula density estimation, we focus on a finite-dimensional marginal density  $c_{1:K}$ corresponding to the first $K$ variables. It is important to note that the consistency and asymptotic properties of the copula density estimator are derived under the condition that $K < \infty$ is fixed and $d \to \infty$. This approach ensures that the latent variable estimation benefits from the complete dataset, while the copula density estimation is confined to the selected $K$-dimensional marginal. 
\end{remark}

\begin{remark} 

The consistency of the proposed estimator is achieved if $n \to \infty$ and $\ln^5n/d \to 0$ which requires a large $d$. In many practical applications,  the number of observed variables is often much smaller than the sample size, but we show in the next section that the proposed approach yields accurate estimates of the copula density even if $d$ is not very large.
\end{remark}

\section{Simulation Study}
This section illustrates the finite sample performance of the proposed kernel estimator of a one-factor copula density, highlighting its advantages over traditional kernel-based methods. Section \eqref{sub3.1} explores techniques of numerical implementation for efficiently computing the proposed model estimator, with the simulation results detailed in Section~\ref{sub3.2}.

\subsection{Simulation implementation}\label{sub3.1}
We conducted the analysis using the R 4.2.3 statistical computing environment \cite{RCoreTeam2023}. The estimation approach followed the procedures outlined in Section \ref{a}. The marginal distributions are estimated using the standard kernel density estimator \eqref{standard} with bandwidth selection using the methods implemented in the \textbf{ks} package (function \textbf{hpi}) and the guidelines provided by \cite{Duong2014}. Subsequently, the linking copula densities are estimated using transformation estimators \eqref{ctilde}, \eqref{cbar}, and \eqref{cjhat}, inspired by the \textbf{kdecopula} package proposed by \cite{Nagler2016a}.

In the final step, we used numerical integration to compute the one-factor copula density estimator as given in \eqref{chat}. For the classical kernel estimator of the one-factor copula density (denoted as $\hat{c}_{mvcde}$, referred to as the naive estimator), we utilized the \textbf{kde} function from the \textbf{ks} package \cite{Duong2014}. This estimator was applied directly to the observed data without accounting for the factor structure. Bandwidth selection for this approach followed the plug-in method by \cite{Chac2010}.

To evaluate the performance of our estimator $\hat{c}$, defined in \eqref{chat}, and the naive estimator $\hat{c}_{mvcde}$, we employed four metrics: root mean squared error (RMSE), mean absolute error (MAE), standard deviation (SD), and bias. We considered the sample sizes $n = 100, 500, 1000$ and $d = 20, 50, 100$ of observed variables used to compute the proxy to the unobserved factor. The analysis proceeded as follows:

\begin{enumerate}
\item Simulate a sample of size $n$ from a $d$-dimensional one-factor copula model \eqref{fcd}, considering two scenarios for all bivariate linking copulas:
\begin{itemize}
\item Gumbel copula with parameter $1.4$.
\item Clayton copula with parameter $2$.
\end{itemize}
\item Estimate the latent variable based on the $d$ observed variables. Use this estimate to compute all the bivariate linking copulas in \eqref{chat} and the copula density linking the first $K = 5, 10$ variables. Numerical integration was then applied to compute the one-factor copula density \eqref{chat}. Results are compared with the naive estimator $\hat{c}_{mvcde}$ obtained via the \textbf{kde} package.
\item Repeat the above steps 1000 times, calculating the RMSE, MAE, SD, and bias for both estimators across 1000 simulations.
\end{enumerate}

\subsection{Simulation results}\label{sub3.2}
Tables \ref{tab1}-\ref{tab4} show the simulation results. It is seen that the RMSEs and the other three performance metrics improve for the proposed estimator for larger values of $d$ and $n$. It is also evident that the proposed estimator significantly outperforms the naive estimator, demonstrating its superior accuracy in estimating the one-factor copula density.
\setlength{\belowcaptionskip}{10pt}
\begin{table}[H]
\centering
\caption{RMSE, MAE, SD, and BIAS computed for our proposed estimator and naive estimator, using the Gumbel linking copulas and $K=5$}
\label{tab1}
\resizebox{0.85\textwidth}{!}{%
\begin{tabular}{c c c c c c c}
\hline
&Estimator& $d$&RMSE&MAE&SD&BIAS\\
\hline
\multirow{4}{*}{$n=100$}&Naive&&0.494&0.866&0.245&$-0.328$\\
&One-Factor&$20$&0.152&0.119&0.088&$-0.042$\\
&One-Factor&$50$&0.139&0.108&0.083&$-0.045$\\
&One-Factor&$100$&0.134&0.105&0.079&$-0.046$\\
\hline
\multirow{4}{*}{$n=500$}&Naive&&0.365&0.322&0.147&$-0.326$\\
&One-Factor&$20$&0.074&0.058&0.041&$-0.036$\\
&One-Factor&$50$&0.070&0.056&0.038&$-0.037$\\
&One-Factor&$100$&0.069&0.055&0.038&$-0.036$\\
\hline
\multirow{4}{*}{$n=1000$}&Naive&&0.349&0.306&0.112&$-0.330$\\
&One-Factor&$20$&0.059&0.047&0.033&$-0.036$\\
&One-Factor&$50$&0.057&0.046&0.032&$-0.036$\\
&One-Factor&$100$&0.054&0.044&0.030&$-0.035$\\
\hline

\end{tabular}
}
\end{table}
\setlength{\belowcaptionskip}{10pt}
\begin{table}[H]
\centering
\caption{RMSE, MAE, SD, and BIAS computed for our proposed estimator and naive estimator, using the Gumbel linking copulas and $K=10$}
\label{tab2}
\resizebox{0.85\textwidth}{!}{%
\begin{tabular}{c c c c c c c }
\hline
&Estimator& $d$ &RMSE&MAE&SD&BIAS\\
\hline
\multirow{4}{*}{$n=100$}&Naive&& 1.793  & 0.984  & 1.29 & $-0.761$\\
&One-Factor&$20$& 0.378 & 0.214 & 0.248  & $-0.055$\\
&One-Factor&$50$& 0.323 & 0.192 & 0.197 & $-0.101$\\
&One-Factor&$100$& 0.320 & 0.190 & 0.182 & $-0.120$ \\
\hline
\multirow{4}{*}{$n=500$}&Naive&&1.047&0.703&0.353&$-0.719$\\
&One-Factor&$20$&0.174&0.109&0.073&$-0.056$\\
&One-Factor&$50$&0.171&0.105&0.071&$-0.092$\\
&One-Factor&$100$&0.164&0.099&0.070&$-0.096$\\
\hline
\multirow{4}{*}{$n=1000$}&Naive&&0.917&0.684&0.286&$-0.712$\\
&One-Factor&$20$&0.121&0.075&0.069&$-0.056$\\
&One-Factor&$50$&0.115&0.064&0.055&$-0.084$\\
&One-Factor&$100$&0.146&0.093&0.056&$-0.099$\\
\hline

\end{tabular}
}
\end{table}
\setlength{\belowcaptionskip}{10pt}
\begin{table}[H]
\centering
\caption{RMSE, MAE, SD, and BIAS computed for our proposed estimator and naive estimator, using the Clayton linking copulas and $K=5$}
\label{tab3}
\resizebox{0.85\textwidth}{!}{%
\begin{tabular}{c c c c c c c}
\hline
&Estimator& $d$ &RMSE&MAE&SD&BIAS\\
\hline
\multirow{4}{*}{$n=100$}&Naive&&0.600&0.468&0.212&$-0.506$\\
&One-Factor&$20$&0.127&0.109&0.084&$-0.032$\\
&One-Factor&$50$&0.121&0.103&0.081&$-0.031$\\
&One-Factor&$100$&0.116&0.101&0.079&$-0.006$\\
\hline
\multirow{4}{*}{$n=500$}&Naive&&0.522&0.430&0.122&$-0.502$\\
&One-Factor&$20$&0.107&0.125&0.094&$-0.021$\\
&One-Factor&$50$&0.106&0.124&0.092&$-0.020$\\
&One-Factor&$100$&0.104&0.121&0.091&$-0.008$\\
\hline
\multirow{4}{*}{$n=1000$}&Naive&&0.431&0.302&0.107&$-0.501$\\
&One-Factor&$20$&0.078&0.061&0.069&$-0.016$\\
&One-Factor&$50$&0.078&0.060&0.070&$-0.016$\\
&One-Factor&$100$&0.073&0.065&0.064&$-0.003$\\
\hline

\end{tabular}
}
\end{table}

\setlength{\belowcaptionskip}{10pt}
\begin{table}[H]
\centering
\caption{RMSE, MAE, SD, and BIAS computed for our proposed estimator and naive estimator, using the Clayton linking copulas and $K=10$}
\label{tab4}
\resizebox{0.85\textwidth}{!}{%
\begin{tabular}{c c c c c c c}
\hline
&Estimator& $d$ &RMSE&MAE&SD&BIAS\\
\hline
\multirow{4}{*}{$n=100$}&Naive&&1.858&0.909&1.157&$-0.583$\\
&One-Factor&$20$&0.478&0.309&0.219&$-0.147$\\
&One-Factor&$50$&0.447&0.291&0.200&$-0.196$\\
&One-Factor&$100$&0.453&0.292&0.207&$-0.173$\\
\hline
\multirow{4}{*}{$n=500$}&Naive&&1.016&0.657&0.464&$-0.580$\\
&One-Factor&$20$&0.296&0.197&0.126&$-0.215$\\
&One-Factor&$50$&0.285&0.190&0.121&$-0.196$\\
&One-Factor&$100$&0.280&0.187&0.119&$-0.196$\\
\hline
\multirow{4}{*}{$n=1000$}&Naive&&0.785&0.560&0.269&$-0.584$\\
&One-Factor&$20$&0.275&0.190&0.106&$-0.231$\\
&One-Factor&$50$&0.261&0.179&0.102&$-0.215$\\
&One-Factor&$100$&0.263&0.182&0.101&$-0.219$\\
\hline

\end{tabular}
}
\end{table}

\section{Empirical study}\label{empirical}
In this section, we apply the proposed methodology to analyze financial data. The dataset comprises S\&P~500 daily stock returns from the Industrials Sector. It consists of a total of $d=41$ stocks with the following tickers: AXON, BA, GE, GD, HWM, HII, LHX, LMT, NOC, RTX, TXT, TDG, DOV, FTV, IEX, HUBB, ITW, NDSN, PH, PNR, SNA, SWK, GWW, XYL, AOS, ALLE, BLDR, JCI, Mas, TT, CAT, CMI, PCAR, WAB, J, PWR, AME, ETN, EMR, GNRC, ROK. The stock returns are observed from January 2, 2017, to July 31, 2024, spanning a sample size of $n=1906$ days. 

To address serial autocorrelation, we employ an \textbf{AR(1)–GARCH(1,1)} model featuring symmetric Student's-$t$ innovations to model univariate marginals for log-returns $\tau_{j,t}=\log (S_{j,t}/S_{j,t-1})$ where $S_{j,t}$ is the $j$th stock price at time $t=1,\ldots,n$: 
\begin{eqnarray*}
    \tau_{j,t} &=& \alpha _{j,0} + \alpha _{j,1} \tau_{j,t-1} + \delta_{j,t} \zeta_{j,t},\\
    \delta_{j,t}^2 &=& \omega_{j,0} + \omega_{j,1} \delta_{j,t-1}^2 \zeta_{j,t-1}^2 + \omega_{j,2} \delta_{j,t-1}^2.
\end{eqnarray*}

The error terms, denoted as $\zeta_{j,t}$, are independently and identically (i.i.d.) $T_{\mu_j}$-distributed with $\mu_j$ degrees of freedom for the $j$th stock. We used the Ljung–Box test to assess whether the fitted residuals are uncorrelated. The model standardized residuals are then transformed to uniform scores $\{ (u_{1, i},\ldots,u_{41, i})\}_{i=1}^n$. 

\begin{figure}[H]
    \centering
        \includegraphics[width=0.45\textwidth]{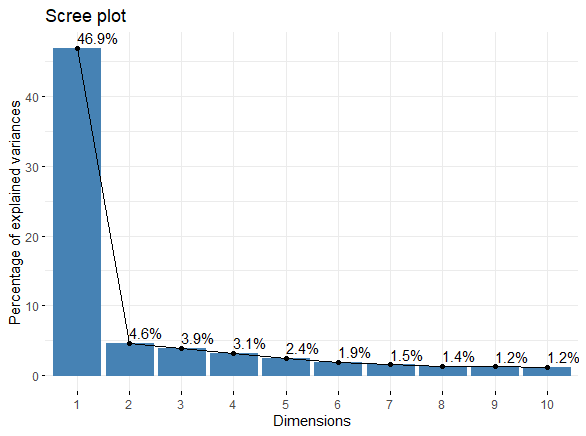}
    \caption{Scree plot of the ten largest eigenvalues of the rank correlation matrix estimated from the AR(1)-GARCH(1,1) filtered stock returns from the Industrials sector of the $S\&P500$ index.}
    \label{scr}
\end{figure}

In the subsequent discussion in Section 2.4 of  \cite{Oh2017} which is inspired by \cite{Cattel1966}, it is shown that, subject to certain regularity conditions, it is possible to determine the optimal number of factors by examining a scree plot based on the eigenvalues of the rank correlation matrix. Figure \eqref{scr} shows the scree plot for the standardized residuals. The results indicate that a single factor effectively accounts for the dominant variance in the data.

\begin{figure}[!h]
    \centering
    
        \includegraphics[width=1\textwidth]{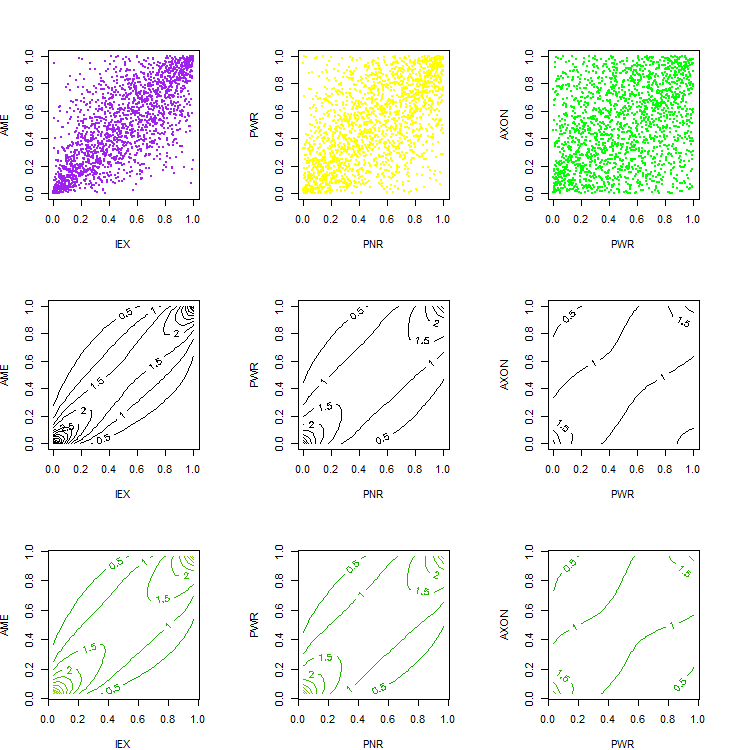}

    \caption{Scatter plots (top row) and estimated contour plots of the bivariate copula densities for some pairs of GARCH-filtered stock returns from the Industrials sector of the S\&P 500 index using the classical copula estimator (middle row) and proposed estimator (bottom row).}
    \label{scat1.1}
\end{figure}

To evaluate the performance of our estimator, we apply the proposed estimator and classical bivariate copula estimator \cite{Nagler2016b} to each pair of the residuals transformed to the $U(0,1)$ scale. We use the bivariate copula estimator as a benchmark since it does not rely on any assumptions (including a factor structure) and performs very well for bivariate data sets. Figure \eqref{scat1.1} shows scatter plots and estimated contour plots of the copula density linking some pairs of variables.
In the figure, the columns (from left to right) show the results for pairs with strong, moderate, and weak dependence, while the rows show scatter plots (top row), the bivariate copula density estimates using the classical copula estimator (middle row), and the proposed one-factor copula density estimator (bottom row).

The results demonstrate that our estimator and the copula estimator yield very similar results for bivariate marginals. To quantify the difference between the two estimators, we use the root mean squared difference (RMSD) for each sample between our proposed estimator and the true one-factor copula model, as well as between the classical bivariate copula density estimator and the true model, as follows:
\begin{eqnarray}\label{RMSD}
     RMSD (\hat c_K) = \left\{\frac{1}{n}\sum_{i=1}^n (\hat{c}(u_{i1},\ldots,u_{iK})-c(u_{i1},\ldots,u_{iK}))^2\right\}^{1/2},
\end{eqnarray}
where $\hat{c}(u_{i1},\ldots,u_{iK})$ is either $\hat{c}_{1:2}(u_{i1},u_{i2})$ (our proposed estimator based on equation \eqref{chat} in bivariate case $K=2$) or $\hat{c}_{kdecop; i,j}(u_i,u_j)$ (the classical bivariate copula estimator in \cite{Nagler2016b}) and 
$c(u_{1},\ldots,u_{K})$ is the true one-factor copula \eqref{fcd} in bivariate case (when the dimension is $2$). In the bivariate case, we consider all $\binom{41}{2} = 820$ unique pairs of stocks from the Industrial Sector, where each pair $(u_i,u_j)$ corresponds to a distinct combination of two variables among the $d=41$ stocks. We compute the RMSD for each of these $820$ pairs. The value of $RMSD(\hat c_K)$, averaged over all pairs $(i,j)$ of stocks,  is 0.08, which shows that the two estimators are almost identical.

\begin{figure}[H]
    \centering
    
        \includegraphics[width=1\textwidth]{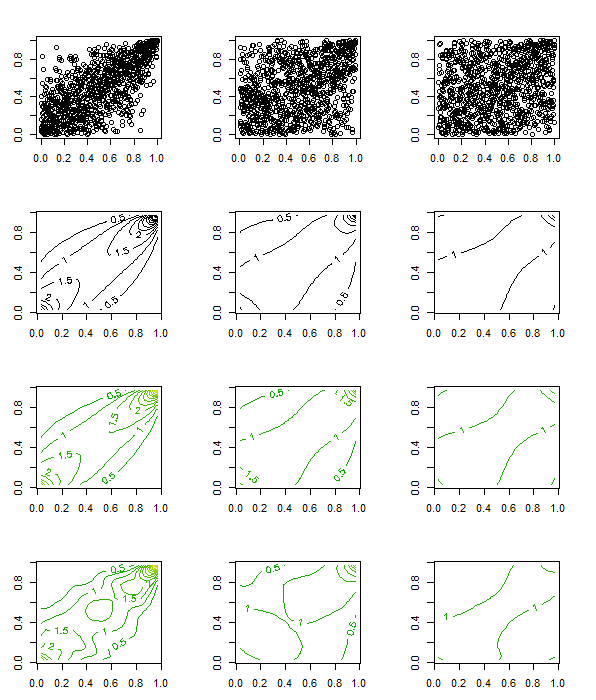}
        
    \caption{Scatter plots for some pairs of simulated variables (first row), the contour plots of the true one-factor copula density (second row), the estimated contour plots using the proposed  estimator (third row), and the estimated contour plots using the classical copula estimator (fourth row)} 
    \label{scat1.2}
\end{figure}

\begin{figure}[H]
    \centering
    
        \includegraphics[width=1\textwidth]{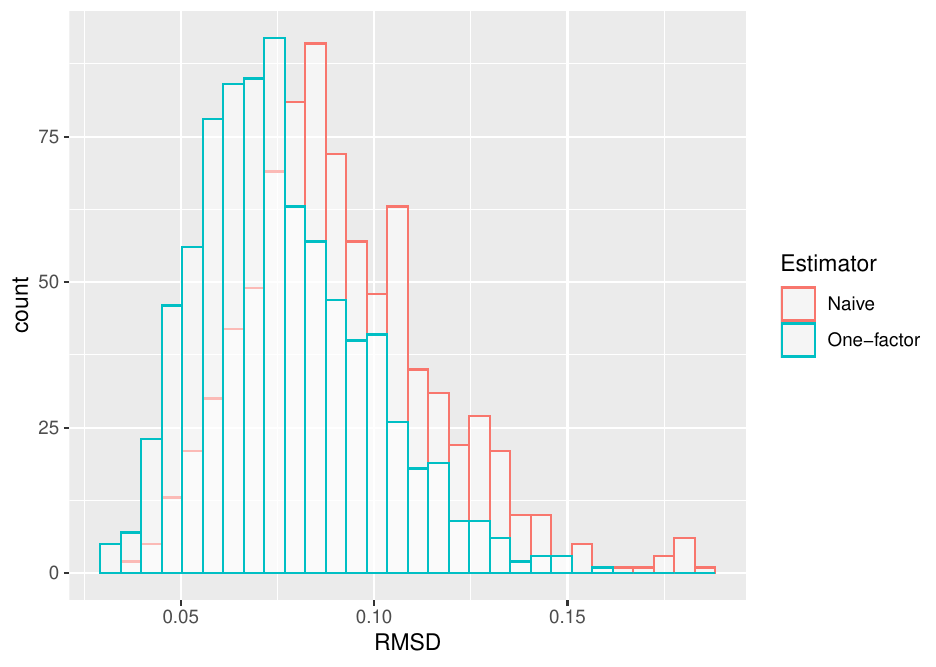}

    \caption{Histogram of RMSDs calculated for different pairs of variables generated from the one-factor copula, obtained using the classical copula density estimator (red) and the proposed estimator (blue).}
    \label{hist1}
\end{figure}

Furthermore, we fitted several parametric one-factor copula models to the dataset comprising 41 stocks from the Industrial Sector. The best-fitting model was identified using the Bayesian Information Criterion (BIC), which selected the model with the Gumbel linking copulas. Assuming this is the true model for the standardized residuals, we generated a synthetic data set of a sample of size 1000 from this model and repeated the above steps for the simulated data set.  Namely, we applied our proposed estimator and the classical copula estimator to each pair of variables from the simulated data set. To evaluate the performance of the two estimators, we constructed contour plots of the estimated copula densities; Figure \eqref{scat1.2} shows results for some pairs of variables, and Figure \eqref{hist1} shows the histogram of RMSDs between each of the two estimators and the true copula density, computed for different pairs of variables from the generated data set using the classical (red) and proposed estimators (blue). 
These results clearly show that our estimator outperforms the classical estimator when the data come from the one-factor model.

To evaluate the performance of our estimator in higher dimensions, we randomly select 1000 triplets from the generated data. We compute the RMSDs for each triplet between our proposed estimator ($\hat{c}_{1:3}(u_{i1},u_{i2},u_{i3})$ defined in \eqref{chat}) and the true one-factor copula model (defined in \eqref{fcd} when the dimension is $3$) as well as between the naive estimator and the true model using equation \eqref{RMSD} when $K = 3$, and  $\hat{c}(u_{i1},u_{i2},u_{i3})$ can be either the proposed estimator or the naive one. 

\begin{figure}[H]
    \centering
    \includegraphics[width=1\textwidth]{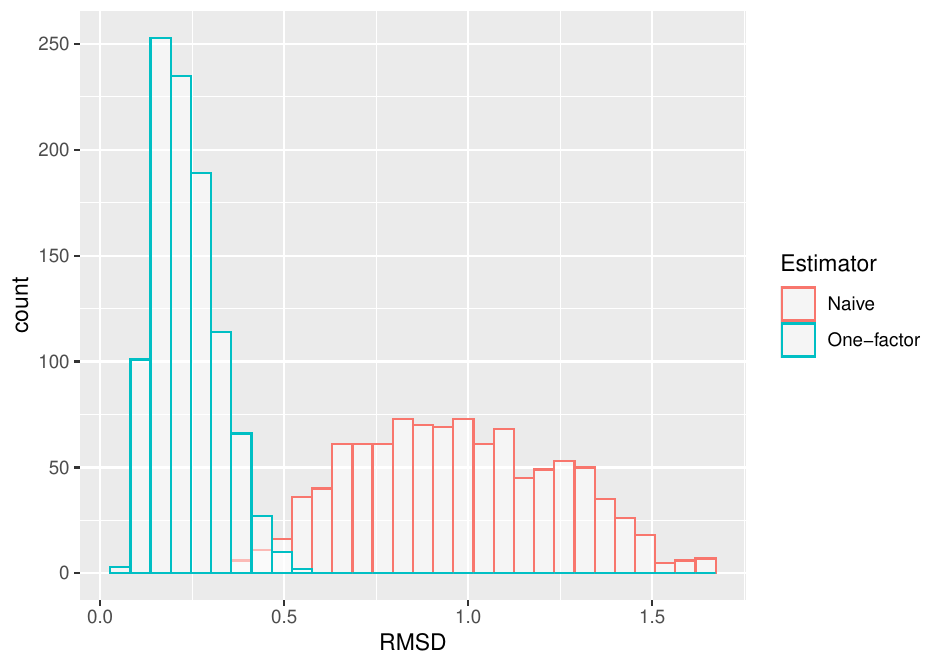}
    \caption{Histogram of RMSDs calculated for different triplets of variables generated from the one-factor copula, obtained using the classical copula density estimator (red) and the proposed estimator (blue).}
    \label{hist2}
\end{figure}

We use the naive estimator, denoted as $\hat{c}_{mvcde}(u_{i1},\ldots,u_{iK})$ introduced in Section \ref{sub3.1}, which differs from the classical bivariate copula estimator, $\hat{c}_{kdecop}$ as used and defined before in the bivariate case. Figure \eqref{hist2} shows the histograms of the RMSDs, which clearly demonstrate that the RMSDs associated with our proposed estimator are significantly smaller than those of the naive estimator. This finding highlights the superior performance of our approach in high-dimensional settings.

Finally, for the copula density linking the first $K=10$ variables, we compute the RMSDs between the best parametric model in terms of BIC and the naive estimator as well as the proposed estimator. We use equation \eqref{RMSD} to compute RMSDs for $K=10$ and different dimensions $d$; table \eqref{tab5} shows the results.

We then repeated the analysis for the data from the Health Care and Energy sectors, with tickers and some results provided in Appendix \ref{Appdx2}. For all three different sectors Table \eqref{tab5} shows that the proposed copula density estimator is much closer to the best parametric copula model. In this table, $d$ represents the number of stocks in each sector.

\setlength{\belowcaptionskip}{10pt}
\begin{table}[h!]
\centering
\caption{RMSDs between the true copula density linking first $K=10$ variables and proposed factor copula density estimator as well as the naive estimator computed using the GARCH-filtered returns from January 2, 2017, to July 31, 2024. }
\label{tab5}
\resizebox{1\textwidth}{!}{%
\begin{tabular}{c c c c c c c c c c }
\hline

\textbf{Sectors}&&&&\textbf{Factor copula estimator}&&\textbf{Naive estimator}&&&$d$\\
\hline
Industrial&&&&0.57&&1.18&&&41\\
Health Care&&&&0.74&&1.08&&&37\\
Energy&&&&1.39&&2.61&&&22\\
\hline
\end{tabular}
}
\end{table}

\section{Discussion}
Factor copula models, studied by \cite{Krupskii2013}, offer a parsimonious framework for capturing diverse dependence structures, including tail dependence and asymmetric relationships. However, real-world data often exhibit dependence structures that cannot be adequately modeled by existing parametric approaches, highlighting the necessity of non-parametric methods.

In this paper, we developed a novel non-parametric estimator for the one-factor copula model that combines proxy to the unobserved factor and the kernel estimation approach. We showed the consistency of the proposed estimator and its good finite-sample performance. 

We applied the new estimator for the analysis of stock returns data and showed that our estimator outperforms the naive multivariate density kernel estimator, particularly in high-dimensional settings, while maintaining competitive performance even in the bivariate case. These findings underscore the potential of the proposed estimator for the  analysis of data sets with complex dependence structures in real-world applications.

There are several potential directions for future research and applications, including:
\begin{itemize}
    \item Extending our approach to different types of dependent variables: this involves adapting the methodology to handle dynamic dependence structures that evolve over time, as well as mixed continuous-ordinal data where variables may have distinct measurement scales. Furthermore, the approach could be expanded to address datasets with missing values by incorporating robust imputation methods and analyzing item response variables, which are commonly used in fields like psychometrics and educational testing.
    \item Developing non-parametric estimation methods for structured factor copula models, including nested, oblique, and $p$-factor copula models, using the proxy method outlined by \cite{Fan2023}.
\end{itemize} 

\vspace{1em}
\noindent\textbf{Acknowledgments} \\[0.5em]
The authors are grateful to the referees and the Associate Editor for their constructive comments, which greatly enhanced the quality of the article.\\

\noindent\textbf{Funding}\\[0.5em]
There is no funding for this paper.\\

\noindent\textbf{Data Availability}\\[0.5em]
The data sets used and analyzed during the current study are available from the corresponding author on reasonable request. 
\section*{Declarations}
\textbf{Conflict of interest}\\
The authors declare no conflict of interest.\\

\section{Appendix}
\subsection{Proofs Theorem \ref{theo1} and Proposition \ref{pro1}}
\label{proof}

This section illustrates the proof of Theorem \eqref{theo1} along with the proof of Proposition \eqref{pro1}. We will use the following notation throughout this section:
 \begin{eqnarray}
  \nonumber Z_j^{(i)} &=& \Phi^{-1} (U_j^{(i)}),~~W^{(i)} = \Phi^{-1} (V_0^{(i)}),\\
  \nonumber \hat{Z}_j^{(i)} &=& \Phi^{-1} (\hat{U}_j^{(i)}),~~\Bar{Z}_d = \frac{1}{d} \sum_{j=1}^d \Phi^{-1} (U_j^{(i)}).
 \end{eqnarray}
 Here, $V_0$ is the latent variable and $\hat{W}=\Phi^{-1}(\hat{F}(\bar{Z}_d))$ is the proxy for $W$, where $\hat{F}$ denotes the CDF of $\Bar{Z}_d$ as before.\\\\
\textbf{Proof of Proposition \eqref{pro1}:} For the first equation in \eqref{pro1.eq}, we employ the proof techniques from \cite{Nagler2017} to analyze the bias and variance of our kernel estimator in the presence of the proxy variable. Suppose that 
\begin{equation}
   \nonumber \hat{f}(\boldsymbol{x}) = \frac{1}{n~det(B_n)} \sum_{i=1}^n K_{B_n} (\boldsymbol{x} - \boldsymbol{X}_i),
\end{equation}
where $K_{B_n}$ is defined in \eqref{cbar}. Then, the bias and variance terms are 
\begin{eqnarray}
  \nonumber Bias[\hat{f}(x,y)] &=& \frac{\sigma^2_K}{2} tr\{(B_nB_n^{\top})Hess~f(x,y)\} + o(det(B_n)), \\
  \nonumber Var[\hat{f}(x,y)] &=& \frac{d^2_K}{n~det(B_n)} f(x,y) + o\left( 
\frac{1}{n~det(B_n)} \right)\,,
\end{eqnarray}
where $\sigma_K^2 = \int_0^1 s^2 K(s) \dd s$ and $d_K = \int_0^1 K^2(s) \dd s$. Also the bandwidth matrix $B_n = \begin{pmatrix}
b_{1n} & 0 \\
b_{3n} & b_{2n} 
\end{pmatrix}$ has three parameters $b_{1n}, b_{3n}, b_{2n}>0$. 
Using the definition of linking copula in \eqref{cbar} and the notation in Section \ref{b}, we have
\begin{equation}
    \nonumber \tilde{c}_{U_j, V_0} (u_j,v_0) = \tilde{c}_{U_j, V_0} (\Phi(z_j),\Phi(w)) = \frac{\hat{f}(z_j,w)}{\phi(z_j)\phi(w)}.
\end{equation}
To proceed, we begin by calculating the variance of the term $\tilde{c}_{U_j, V_0} (\Phi(z_j),\Phi(w))$ as follows
\begin{eqnarray}
   \nonumber Var \left[ \tilde{c}_{U_j, V_0} (\Phi(z_j),\Phi(w)) \right] &=& \frac{1}{\phi^2(z_j)\phi^2(w)} Var[\hat{f}(z_j,w)] \\
   \nonumber &=& \frac{1}{\phi^2(z_j)\phi^2(w)} \left[ \frac{d^2_K}{n~det(B_n)} f(x,y) + o\left( 
\frac{1}{n~det(B_n)} \right) \right] \\
\nonumber &=& \frac{d^2_K}{n~det(B_n)} \frac{c_{U_j, V_0} (\Phi(z_j),\Phi(w))}{\phi(z_j)\phi(w)} + o\left( \frac{1}{n~det(B_n)} \right) \\
\nonumber &=& \frac{d^2_K}{n~det(B_n)} \frac{c_{U_j, V_0} (u_j, v_0)}{\phi(\Phi^{-1}(u_j))\phi(\Phi^{-1}(v_0))} + o\left( \frac{1}{n~det(B_n)} \right).
 \end{eqnarray}
 
The last equality is obtained through a change of variables. To compute the bias, we write
\begin{equation}\label{T}
 \begin{aligned}
     E\left[ \tilde{c}_{U_j, V_0} (\Phi(z_j),\Phi(w)) \right]  = & \ \frac{1}{\phi(z_j)\phi(w)} E[\hat{f}(z_j,w)] \\ 
     = & \  \frac{f(z_j,w)}{\phi(z_j)\phi(w)} + \frac{\sigma^2_K}{2} \frac{tr\{(B_nB_n^{\top})Hess~f(z_j,w)\}}{\phi(z_j)\phi(w)} + o(det(B_n))\\
      = & \ c_{U_j, V_0} (\Phi(z_j),\Phi(w)) + \frac{\sigma^2_K}{2} \frac{tr\{(B_nB_n^{\top})Hess~f(z_j,w)\}}{\phi(z_j)\phi(w)} \\
    & +  o(det(B_n)).
    \end{aligned}
 \end{equation}
By using $f(z_j,w) = c\left( \Phi(z_j), \Phi(w)\right)\phi(z_j)\phi(w)$, $c_u = \frac{\partial}{\partial u} c(u,v)$ and $c_{uu} = \frac{\partial^2}{\partial u^2} c(u,v)$, and noting that $\phi^{'}(x) = -x\phi(x)$, $\phi^{''}(x) = (x^2-1)\phi(x)$, the Hessian matrix of $f(z_j,w)$ can be calculated as follows:
 \begin{eqnarray*}
   \frac{\partial f(z_j,w)}{\partial z_j} &=& \phi (w) \left[ c_u\left( \Phi(z_j), \Phi(w) \right) \phi^2(z_j) - c\left(\Phi(z_j) ,\Phi(w) \right) z_j\phi(z_j)\right],  \\ 
   \frac{\partial^2 f(z_j,w)}{\partial z^2_j} &=& \phi (w) \left[ c_{uu}\left( \Phi(z_j), \Phi(w) \right) \phi^3(z_j) - 3c_u\left( \Phi(z_j), \Phi(w) \right) z_j\phi^2(z_j)\right]\\
   && +\ c\left( \Phi(z_j), \Phi(w) \right) (z^2_j-1)\phi(z_j)\phi(w),  \\ 
   \frac{\partial^2 f(z_j,w)}{\partial z_j \partial w} &=& c\left( \Phi(z_j), \Phi(w) \right) z_j w \phi(z_j) \phi(w) + c_{uv}\left( \Phi(z_j), \Phi(w) \right) \phi^2(z_j) \phi^2(w)\\
   && - \ c_u\left( \Phi(z_j), \Phi(w) \right) \phi^2(z_j) w \phi(w) -  c_v\left( \Phi(z_j), \Phi(w) \right) \phi^2(w) z_j \phi(z_j).
 \end{eqnarray*}
 Let $u_j= \Phi(z_j)$ and $v_0 = \Phi(w)$. We find:
 \begin{eqnarray*}
   T(u_j, v_0) &:=& \frac{tr\{(B_nB_n^{\top})Hess~f(z_j,w)\}}{\phi(z_j)\phi(w)}   \\
   &=& b_{1n}^2 \left[ c_{uu}(u_j,v_0) \phi^2(\Phi^{-1}(u_j)) \right]\\
  && -\  b_{1n}^2 \left[ 3 c_u(u_j,v_0) \Phi^{-1}(u_j) \phi(\Phi^{-1}(u_j))+c(u_j,v_0) (\Phi^{-1}(u_j)^2-1)\right]\\
   && +\  (b_{2n}^2 + b_{3n}^2) \left[ c_{vv}(u_j,v_0) \phi^2(\Phi^{-1}(v_0)) -3 c_v(u_j,v_0) \Phi^{-1}(v_0) \phi(\Phi^{-1}(v_0))\right]  \\
    && + \  (b_{2n}^2 + b_{3n}^2)\left[c(u_j,v_0) (\Phi^{-1}(v_0)^2-1)\right] \\
   && +\  2b_{1n} b_{3n}\left[ c(u_j,v_0) \Phi^{-1}(u_j) \Phi^{-1}(v_0) - c_u(u_j,v_0) \Phi^{-1}(v_0) \phi(\Phi^{-1}(u_j))\right] \\
   && - \  2b_{1n} b_{3n}\left[c_v(u_j,v_0) \Phi^{-1}(u_j)\phi(\Phi^{-1}(v_0)) -c_{uv}(u_j,v_0) \phi(\Phi^{-1}(u_j)) \phi(\Phi^{-1}(v_0))\right].
 \end{eqnarray*}
 From \eqref{T}, the bias term for $\tilde{c}_{j,0} (\Phi(z_j),\Phi(w))$ is
 \begin{equation}
  Bias[\tilde{c}_{U_j, V_0} (\Phi(z_j),\Phi(w))] = \frac{\sigma^2_K}{2} T\left(u_j, v_0 \right) + o(det(B_n)).    
 \end{equation}
 The proof of the first part in Proposition \eqref{pro1} follows directly from Markov's inequality. 
 Now let $\delta(z_j,w) = \Bar{c}(\Phi(z_j),\Phi(w)) - \Tilde{c}(\Phi(z_j),\Phi(w))$ and also define $\phi_n(z_j, w) = n\phi(\Phi^{-1}(u_j))\phi(\Phi^{-1}(v_0))$. To prove the second part of Proposition \eqref{pro1}, we have
\begin{align*}
   \delta(z_j, w) = & \phi_n(z_j, w) \sum_{i=1}^n \boldsymbol{K}_{B_n} \left( \Phi^{-1}(u_j) - \Phi^{-1}(U_j^{(i)}) , \Phi^{-1}(v_0) - \hat{W}^{(i)} \right) \\
    &- \phi_n(z_j, w) \sum_{i=1}^n\boldsymbol{K}_{B_n} \left( \Phi^{-1}(u_j) - \Phi^{-1}(U_j^{(i)}) , \Phi^{-1}(v_0) - W^{(i)} \right) \\
    = & \phi_n(z_j, w) \sum_{i=1}^n K_{B_n} \left[\Phi^{-1}(u_j) - \Phi^{-1}(U_j^{(i)})\right] K_{B_n}\left[\Phi^{-1}(v_0) - \hat{W}^{(i)}\right]\\
    &-\phi_n(z_j, w) \sum_{i=1}^n K_{B_n} \left[\Phi^{-1}(u_j) - \Phi^{-1}(U_j^{(i)})\right]K_{B_n}\left[\Phi^{-1}(v_0) - W^{(i)}\right]. 
\end{align*}
By applying the first-order Taylor approximation to the function $K_{B_n}$, we have: 
\begin{equation}
\label{Op}
\begin{aligned}
    \delta(z_j, w) = & \ \phi_n(z_j, w) \sum_{i=1}^n K_{B_n} (\Phi^{-1}(u_j) - \Phi^{-1}(U_j^{(i)}))\\
      & \times \left[ \left(K_{B_n}(\Phi^{-1}(v_0) - \hat{W}^{(i)})\right)^{'} (\hat{W}-W)+ o(\hat{W}-W) \right]\\
     \leq & \ \left[ \phi_n(z_j, w) \sum_{i=1}^n K_{B_n} (\Phi^{-1}(u_j) - \Phi^{-1}(U_j^{(i)})\right]\\ 
     & \times \left[ \left(K_{B_n}(\Phi^{-1}(v_0) - \hat{W}^{(i)})\right)^{'} \frac{1}{\phi(W)}+ o(\hat{W}-W)\right]\\
    = &\  O_p\left(\frac{\sqrt{\ln n}}{\sqrt{d}}\right) + O_p\left(\frac{1}{\sqrt{ n}}\right).
    \end{aligned}
\end{equation}
 The inequality in \eqref{Op} follows from part (b) of  Assumption \eqref{ass}. Next, we prove the third part of Proposition \eqref{pro1}. We use the first-order Taylor approximation of $\Phi^{-1}$:
 \begin{equation}\label{O} \left(\hat{Z}_j^{(i)} - Z_j^{(i)} \right) = \frac{\left( \hat{U}_j^{(i)} - U_j^{(i)} \right)}{\phi(Z_j^{(i)})} + o\left( \hat{U}_j^{(i)} - U_j^{(i)} \right) 
   = \left[ \frac{1}{\phi(Z_j^{(i)})}\right]\times O_{a.s.}(b_{e,n}).
 \end{equation}

Denote $\xi_z = (\frac{\partial}{\partial z_j})^{'}$. We use the first-order Taylor approximation of $K_{B_n}$ to write:
{\small \begin{align}\label{xi}
  \nonumber \tilde\delta_n(z_j,w)  & :=  \phi(z_j) \phi(w) \left|\hat{c}_{U_j, V_0} (\Phi(z_j),\Phi(w)) - \Bar{c}_{U_j, V_0} (\Phi(z_j),\Phi(w))\right| \\
   \nonumber &=\  \left| \frac{1}{n} \sum_{i=1}^n K_{B_n}(z_j - \hat{Z}_j^{(i)}) K_{B_n}(w - \hat{W}^{(i)}) - \frac{1}{n} \sum_{i=1}^n K_{B_n}(z_j - Z_j^{(i)}) K_{B_n}(w - \hat{W}^{(i)})\right| \\
   \nonumber &= \ \left| \frac{1}{n} \sum_{i=1}^n K_{B_n}(w - \hat{W}^{(i)}) \xi_z \left\{ K_{B_n}(z_j - Z_j^{(i)}) \right\}\left( \hat{Z}_j^{(i)} - Z_j^{(i)}\right) + o_{a.s.}\left( \hat{Z}_j^{(i)} - Z_j^{(i)}\right) \right| \\
 & \leq \ \left| \frac{1}{n} \sum_{i=1}^n K_{B_n}(w - \hat{W}^{(i)}) \xi_z \left\{ K_{B_n}(z_j - Z_j^{(i)}) \right\} \left( \frac{1}{\phi(Z_j^{(i)})}\right) \right| \times O_{a.s.} (b_{e,n}).
 \end{align}}
Let $\eta_n(z) = \sup_{x \in [z-B_n, z+ B_n]} \frac{1}{\phi(x)} = O(1)$,
we find from \eqref{xi} that
 \begin{equation}
  \tilde\delta_n(z_j,w) \leq \left[ | K_{B_n}(w - \hat{W}^{(i)})\xi_z \theta(z)| + o_{a.s.}(1) \right] \times O_{a.s.} (b_{e,n}).
 \end{equation}
where the derivative of $\theta (z_j):= K_{B_n}(z_j - Z_j^{(i)})$ is continuous and bounded by Assumption \eqref{ass}, so we can use the boundedness of the kernel function $K_{B_n}$ to complete the proof. \hfill $\Box$

\bigskip
\textbf{Proof of Theorem \eqref{theo1}:}
To obtain the result, first, we consider the rate of convergence of the linking copula estimator:
 \begin{multline*}
     |\hat{c}_{U_j, V_0} (u_j,v_0) - c_{U_j, V_0} (u_j,v_0)|
   \leq |\hat{c}_{U_j, V_0} (u_j,v_0) - \Bar{c}_{U_j, V_0} (u_j,v_0)|\\ + |\Bar{c}_{U_j, V_0} (u_j,v_0) - \tilde{c}_{U_j, V_0} (u_j,v_0)|
  +\ |\tilde{c}_{U_j, V_0} (u_j,v_0) - c_{U_j, V_0} (u_j,v_0)|\\
    = A_1+A_2+A_3.
 \end{multline*}
 Recall, $b_{e,n} = \sup_{i=1,\ldots,n} |\hat{U}_j^{(i)} - U_j^{(i)}|$ and by Assumption \eqref{ass}, part (a), one can obtain
 \begin{eqnarray}\label{a1.1}
  |\hat{U}_j^{(i)} - U_j^{(i)}| = |\hat{\Phi}(Z_j) - \Phi(Z_j)| \leq \sup_{z_j}|\hat{F}(z_j) - F(z_j)| = o_{a.s.}(n^{-r}).   
 \end{eqnarray}
Without loss of generality,  we assume that $B_n = b_n \times I_2$, where $I_2$ is the identity matrix. When using the mean-squared optimal bandwidth $b_n = O_p(n^{-\frac{1}{6}})$ and $p=2$, we obtain $A_3 = O_p(n^{-r})$. So, using Proposition \eqref{pro1} for $A_1$ and $A_2$, we have
\begin{eqnarray}
  \hat{c}_{U_j, V_0} (u_j,v_0) - c_{U_j, V_0} (u_j,v_0) = O_p(n^{-s}) = O_p(n^{-r}) + O_p((\ln n/  d)^{1/2}) + O_p(n^{-1/2}).  
\end{eqnarray}

To complete the proof of Theorem \eqref{theo1}, we will use the Dominated Convergence Theorem (DCT). We begin by demonstrating that the estimator $\hat{c}_{U_j, V_0}$ is bounded. We assume that $0 < u_{\min} < u_j < u_{\max} < 1$ for any $j = 1, \ldots, K$, and further assume that there exists a constant $K_0 > 0$ such that $c_{U_j, V_0}(u_j, v_0) < K_0$ for all $u_{\min} < u_j < u_{\max}$ and $0 \leq v_0 \leq 1$. These are mild assumptions that are satisfied by many commonly used parametric copula families provided that dependence is not very strong.

We can redefine $\hat{c}_{U_j, V_0}(u_j, v_0) := \min\{\hat{c}_{U_j, V_0}(u_j, v_0), K_0\}$ which does not affect the asymptotic behavior of the estimator. Using the boundedness of the estimator $\hat c_{U_j, V_0}$, we apply CDT to obtain
\begin{eqnarray*}
    \hat{c}_{1:K}(u_1,\ldots,u_k) &=& \int_0^1 \prod _{j=1}^K \hat{c}_{U_j, V_0} (u_j,v_0) \dd v_0 
    = \int_0^1 \prod _{j=1}^K c_{U_j, V_0} (u_j,v_0)\dd v_0 + O_p(n^{-s})
    \\ &=& c(u_1,\ldots,u_k) + O_p(n^{-s}),
\end{eqnarray*}
which concludes the proof. \hfill $\Box$

\subsection{Analysis of data from the Energy and Health Care sectors}
\label{Appdx2}

We conducted a similar analysis of the stocks from the Energy and Health Care sectors. For the Energy sector, we considered $d=22$ stocks with the following tickers: CVX, XOM, HES, BKR, HAL, SLB, APA, COP, CTRA, DVN, FANG, EOG, EQT, MRO, OXY, MPC, PSX, VLO, KMI, OKE, TRGP, WMB. The stock returns are observed from January 2, 2017, to July 31, 2024, spanning a sample size of $n=1906$ days. For the Health Care sector (encompassing Biotechnology, Services, Equipment, and Distributors industries), we considered $d=37$ stocks with the following tickers: AMGN, ABBV, BIIB, GILD, INCY, REGN, VRTX, ABT, BAX, BDX, DXCM, BSX, PODD, EW, HOLX, IDXX, ZBH, ISRG, MDT, RMD, RVTY, STE, SYK, TFX, CAH, COR, HSIC, MCK, A, BIO, TECH, CRL, DHR, IQV, MTD, TMO, WAT. The stock returns are observed from January 2, 2017, to December 30, 2020, spanning a sample size of $n=1006$ days. This shorter period was chosen to ensure that the joint dependence of data from the Health Care sector can be adequately captured by a single common factor. When using a longer time frame, the data indicated the presence of two common factors, likely due to structural shifts in the market, particularly the impact of the COVID-19 pandemic. To maintain consistency with the one-factor assumption, we restricted the analysis to this period.

\begin{figure}[H]
    \centering
    \subfigure[Health Care sector data]{
        \includegraphics[width=0.5\textwidth]{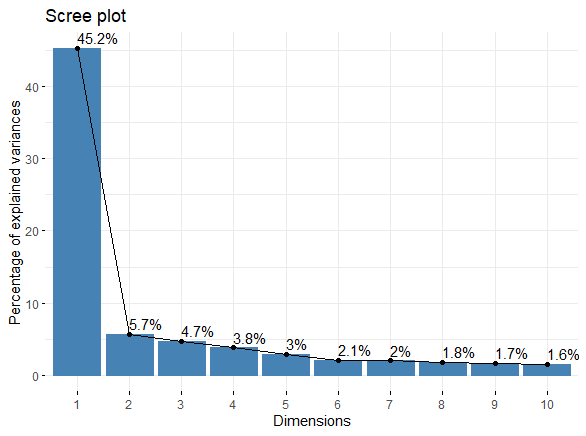}
    }
    \hfill 
    \subfigure[Energy sector data]{
        \includegraphics[width=0.45\textwidth]{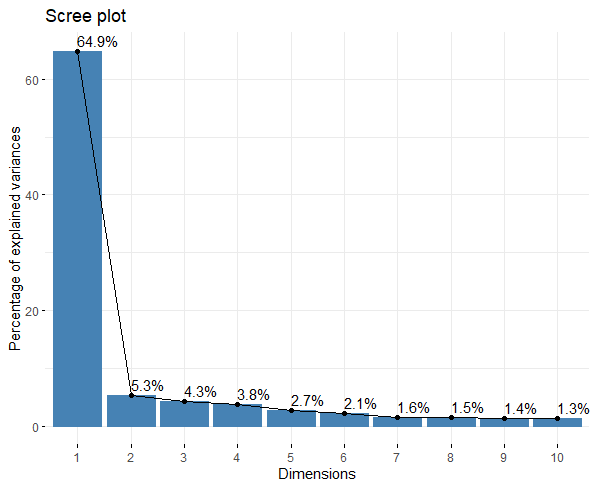}
    }
    \caption{Scree plot of the ten largest eigenvalues of the rank correlation matrix estimated from the AR(1)-GARCH(1,1) filtered stock returns from the Health Care and Energy sectors of the $S\&P500$ index.}
    \label{scr.2}
\end{figure}

Similar to the Industrials sector data, we applied the \textbf{AR(1)--GARCH(1,1)} model to the returns from the two sectors, and the scree plot indicates a one-factor structure is suitable for the GARCH-filtered stock returns from these sectors; see Fig. \eqref{scr.2}.  

\begin{figure}[H]
    \centering
        \includegraphics[width=1.15\textwidth]{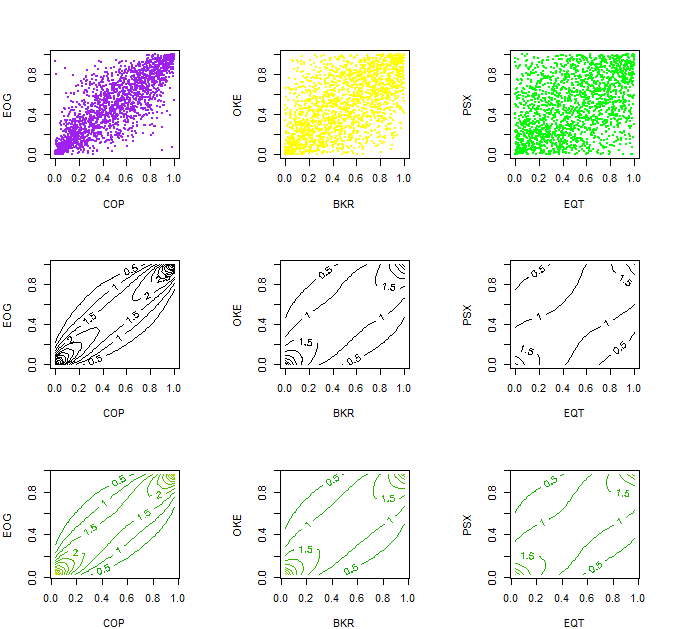}

    \caption{Scatter plots (top row) and estimated contour plots of the bivariate copula densities for some pairs of GARCH-filtered stock returns from the Energy sector of the $S\&P500$ index using the classical copula estimator (middle row) and proposed estimator (bottom row).}
    \label{scat1.3}
\end{figure}

\begin{figure}[H]
    \centering

    \includegraphics[width=1.15\textwidth]{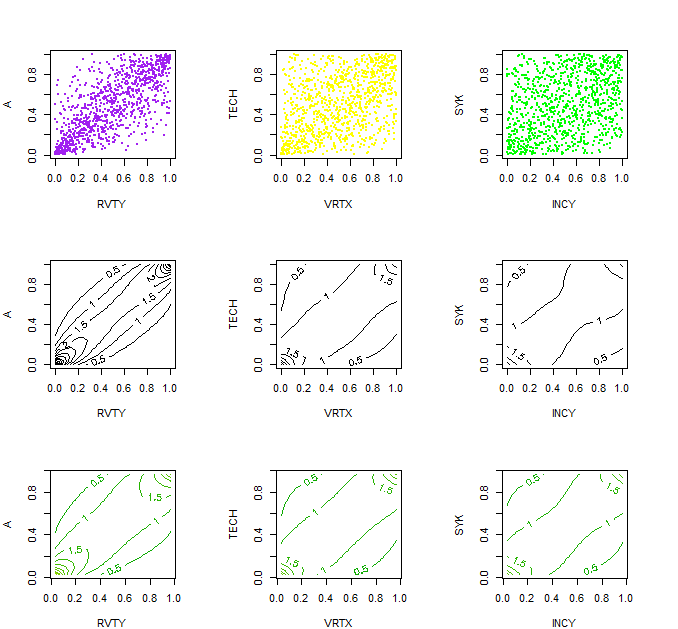}

    \caption{ Scatter plots (top row) and estimated contour plots of the bivariate copula
densities for some pairs of GARCH-filtered stock returns from the Health Care sector of the $S\&P 500$ index using the classical copula estimator (middle row) and proposed estimator
(bottom row).}
    \label{scat1.4}
\end{figure}

Finally, figures \eqref{scat1.3} and \eqref{scat1.4} show the  estimated contour plots of the copula density linking some pairs of variables from the two sectors.  Again, the results demonstrate that the proposed estimator and the classical copula estimator yield very similar results for bivariate marginals.

\bibliographystyle{apalike}
\bibliography{mybibliography}

\begin{thebibliography}{}

\bibitem[Autin et~al., 2010]{Autin2010}
Autin, F., Pennec, E.~L., and Tribouley, K. ((2010)).
\newblock Thresholding methods to estimate the copula density.
\newblock {\em Journal of Multivariate Analysis}, 101(1):200--222.

\bibitem[Bellman, 2003]{Bellman2003}
Bellman, R.~E. ((2003)).
\newblock Dynamic programming.
\newblock {\em New York, NY, USA: Dover}.

\bibitem[Bouezmarni et~al., 2013]{Bou2013}
Bouezmarni, T., El~Gouch, A., and Taamouti, A. ((2013)).
\newblock Bernstein estimator for unbounded copula densities.
\newblock {\em Statistics and risk modeling}, 30(4):343–360.

\bibitem[Bouezmarni et~al., 2010]{Bou2010}
Bouezmarni, T., Rombouts, J., and Taamouti, A. ((2010)).
\newblock Asymptotic properties of the bernstein density copula estimator for
  $\alpha$-mixing data.
\newblock {\em Journal of Multivariate Analysis}, 101(1):1--10.

\bibitem[Brechmann et~al., 2012]{Brechmann2012}
Brechmann, E., Czado, C., and Aas, K. ((2012)).
\newblock Truncated regular vines in high dimensions with applications to
  financial data.
\newblock {\em Canadian Journal of Statistics}, 40:68–85.

\bibitem[Cattell, 1966]{Cattel1966}
Cattell, R. ((1966)).
\newblock The scree test for the number of factors.
\newblock {\em Multivariate Behav Res}, 1(2):245--276.

\bibitem[Chac´on and Duong, 2010]{Chac2010}
Chac´on, J. and Duong, T. ((2010)).
\newblock Multivariate plug-in bandwidth selection with unconstrained pilot
  bandwidth matrices.
\newblock {\em TEST}, 19:375–398.

\bibitem[Chatelain et~al., 2020]{Cha2020}
Chatelain, S., Fougères, A., and Nešlehová, J. ((2020)).
\newblock Inference for archimax copulas.
\newblock {\em The Annals of Statistics}, 48(2):1025--1051.

\bibitem[Chatrabgoun et~al., 2017]{Chat2017}
Chatrabgoun, O., Parham, G., and Chinipardaz, R. ((2017)).
\newblock A legendre multiwavelets approach to copula density estimation.
\newblock {\em Stat. Pap}, 58(1):673–690.

\bibitem[Chen et~al., 2015]{Chen2015}
Chen, H., MacMinn, R., and Sun, T. ((2015)).
\newblock Multi-population mortality models: A factor copula approach.
\newblock {\em Insurance: Mathematics and Economics}, 63:135--146.

\bibitem[Chernozhukov et~al., 2017]{Chernozhukov2017}
Chernozhukov, V., Chetverikov, D., and Kato, K. ((2017)).
\newblock Central limit theorems and bootstrap in high dimensions.
\newblock {\em The Annals of Probability}, 45(4):2309--2352.

\bibitem[Duong, 2014]{Duong2014}
Duong, T. ((2014)).
\newblock ks: Kernel smoothing. r package version 1.9.3..

\bibitem[Fan and Joe, 2023]{Fan2023}
Fan, X. and Joe, H. ((2023)).
\newblock High-dimensional factor copula models with estimation of latent
  variables.
\newblock {\em Journal of Multivariate Analysis}, 201:1--29.

\bibitem[Genest et~al., 2009]{Gen2009}
Genest, C., Masiello, E., and Tribouley, K. ((2009)).
\newblock Estimating copula densities through wavelets.
\newblock {\em Insur. Math. Econ}, 44:p.e1557.

\bibitem[Ghanbari and Shirazi, 2023]{Ghanbari2023}
Ghanbari, B. and Shirazi, E. ((2023)).
\newblock Using copula information in wavelet estimation of bivariate density
  function based on censorship observations.
\newblock {\em Communications in Statistics - Theory and Methods},
  53(5):1810--1824.

\bibitem[Ghanbari et~al., 2019]{Ghanbari2019}
Ghanbari, B., Yarmohammadi, M., Hosseinioun, N., and Shirazi, E. ((2019)).
\newblock Wavelet estimation of copula function based on censored data.
\newblock {\em Journal of Inequalities and Applications}, 2019(1).

\bibitem[He et~al., 2021]{Ye2021}
He, Y., Ye, X., Huang, D., Huang, J.~Z., and Zhai, J.-H. ((2021)).
\newblock Novel kernel density estimator based on ensemble unbiased
  cross-validation.
\newblock {\em Information Sciences}, 581:327--344.

\bibitem[Hull and White, 2004]{Hull2004}
Hull, J. and White, A. ((2004)).
\newblock Valuation of a cdo and an nth to default cds without monte carlo
  simulation.
\newblock {\em J. Derivatives}, 12:8--23.

\bibitem[Jin et~al., 2021]{Jin2021}
Jin, Y., He, Y., and Huang, D. ((2021)).
\newblock An improved variable kernel density estimator based on $l_2$
  regularization.
\newblock {\em Mathematics}, 9(16).

\bibitem[Kauermann et~al., 2013]{Kau2013}
Kauermann, G., Schellhase, C., and Ruppert, D. ((2013)).
\newblock Flexible copula density estimation with penalized hierarchical
  b-splines.
\newblock {\em Scandinavian Journal of Statistics}, 40(4):685–705.

\bibitem[Kiriliouk et~al., 2018]{kiriliouk2018estimator}
Kiriliouk, A., Segers, J., and Tafakori, L. (2018).
\newblock An estimator of the stable tail dependence function based on the
  empirical beta copula.
\newblock {\em Extremes}, 21:581--600.

\bibitem[Kirkby et~al., 2023]{Kirkby2023}
Kirkby, J., Leitao, A., and Nguyen, D. (2023).
\newblock Spline local basis methods for nonparametric density estimation.
\newblock {\em Statistics Surveys}, 17:75--118.

\bibitem[Ko and Hjort, 2019]{Ko2019}
Ko, V. and Hjort, N. ((2019)).
\newblock Model robust inference with two-stage maximum likelihood estimation
  for copulas.
\newblock {\em Journal of Multivariate Analysis}, 171:362--381.

\bibitem[Koike, 2021]{Koike2021}
Koike, Y. ((2021)).
\newblock Notes on the dimension dependence in high-dimensional central limit
  theorems for hyperrectangles.
\newblock {\em Japanese Journal of Statistics and Data Science}, 4:257--297.

\bibitem[Krupskii and Genton, 2017]{Krupskii2017}
Krupskii, P. and Genton, M. ((2017)).
\newblock Factor copula models for data with spatio-temporal dependence.
\newblock {\em Spatial Statistics}, 22(1):180--195.

\bibitem[Krupskii et~al., 2018]{Krupskii2018}
Krupskii, P., Huser, R., and Genton, M. ((2018)).
\newblock Factor copula models for replicated spatial data.
\newblock {\em Journal of the American Statistical Association},
  113(521):467--479.

\bibitem[Krupskii and Joe, 2013]{Krupskii2013}
Krupskii, P. and Joe, H. ((2013)).
\newblock Factor copula models for multivariate data.
\newblock {\em J. Multivariate Anal}, 120:85--101.

\bibitem[Krupskii and Joe, 2015]{Krupskii2015}
Krupskii, P. and Joe, H. ((2015)).
\newblock Structured factor copula models: Theory, inference and computation.
\newblock {\em Journal of Multivariate Analysis}, 138:53--73.

\bibitem[Krupskii and Joe, 2022]{Krupskii2022}
Krupskii, P. and Joe, H. ((2022)).
\newblock Approximate likelihood with proxy variables for parameter estimation
  in high-dimensional factor copula models.
\newblock {\em Statist Papers}, 63(2):543--569.

\bibitem[Liu et~al., 2024]{Liu2024}
Liu, H., Havrilla, A., Lai, R., and Liao, W. ((2024)).
\newblock Deep nonparametric estimation of intrinsic data structures by chart
  autoencoders: Generalization error and robustness.
\newblock {\em Applied and Computational Harmonic Analysis}, 68.

\bibitem[Liu et~al., 2021]{Liu2021}
Liu, W., Liang, S., and Qin, X. ((2021)).
\newblock A novel dimension reduction algorithm based on weighted kernel
  principal analysis for gene expression data.
\newblock {\em PLoS One}, 16(10):e0258326.

\bibitem[Modak, 2023]{Modak2023}
Modak, S. ((2023)).
\newblock A new measure for the assessment of clustering based on kernel
  density estimation.
\newblock {\em Communications in Statistics - Theory and Methods},
  52(17):5942--5951.

\bibitem[Nagler, 2016]{Nagler2016a}
Nagler, T. ((2016)).
\newblock kdecopula: Kernel smoothing for bivariate copula densities. r package
  version 0.2.1, url: https://github.com/tnagler/ kdecopula.

\bibitem[Nagler, 016b]{Nagler2016b}
Nagler, T. ((2016b)).
\newblock kdecopula: Kernel smoothing for bivariate copula densities.
\newblock {\em R package version 0.8.0.}

\bibitem[Nagler and Czado, 2016]{Nagler2016}
Nagler, T. and Czado, C. ((2016)).
\newblock Evading the curse of dimensionality in nonparametric density
  estimation with simplified vine copulas.
\newblock {\em J. Multivar. Anal}, 151:69--89.

\bibitem[Nagler et~al., 2017]{Nagler2017}
Nagler, T., Schellhase, C., and Czado, C. ((2017)).
\newblock Nonparametric estimation of simplified vine copula models: comparison
  of methods.
\newblock {\em Dependence Modeling}, 5:99--120.

\bibitem[Nguyen et~al., 2019]{Nguyen2019}
Nguyen, H., Ausin, M.~C., and Galeano, P. ((2019)).
\newblock Parallel bayesian inference for high-dimensional dynamic factor
  copulas.
\newblock {\em Journal of Financial Econometrics, Oxford University Press},
  17(1):118--151.

\bibitem[Nikoloulopoulos and Joe, 2015]{Nikoloulopoulos2015}
Nikoloulopoulos, A. and Joe, H. ((2015)).
\newblock Factor copula models for item response data.
\newblock {\em Psychometrika}, 80:126--150.

\bibitem[Oh and Patton, 2017]{Oh2017}
Oh, D. and Patton, A. ((2017)).
\newblock Modeling dependence in high dimensions with factor copulas.
\newblock {\em Journal of Business and Economic Statistics}, 35(1):139--154.

\bibitem[Qu and Yin, 2012]{Qu2012}
Qu, L. and Yin, W. ((2012)).
\newblock Copula density estimation by total variation penalized likelihood
  with linear equality constraints.
\newblock {\em Computational Statistics and Data Analysis}, 56(2):384--398.

\bibitem[{R Core Team}, 2023]{RCoreTeam2023}
{R Core Team} (2023).
\newblock {\em R: A Language and Environment for Statistical Computing}.
\newblock R Foundation for Statistical Computing, Vienna, Austria.

\bibitem[Scott, 1992]{Scott1992}
Scott, D.~W. ((1992)).
\newblock Multivariate density estimation: Theory, practice, and visualization.
\newblock {\em Hoboken, NJ, USA: Wiley}.

\bibitem[Sklar, 1959]{Sklar1959}
Sklar, A. ((1959)).
\newblock Fonctions de répartition à n dimensions et leurs marges.
\newblock {\em Publ. Inst. Stat. Univ. Paris}, 8:229–231.

\bibitem[Sánchez, 2003]{Sanchez2003}
Sánchez, V. ((2003)).
\newblock Advanced support vector machines and kernel methods.
\newblock {\em Neurocomputing}, 55:5--20.

\bibitem[Wang et~al., 2023]{Wang2023}
Wang, W., Wang, B., Chau, K., and Xu, D. ((2023)).
\newblock Monthly runoff time series interval prediction based on woa-vmd-lstm
  using non-parametric kernel density estimation.
\newblock {\em Earth Science Informatics}, 16:2373--2389.

\end{thebibliography}

\end{document}